\documentclass{elsart}
\usepackage{graphicx}
\usepackage{tabls,tabularx}
\usepackage{longtable}
\usepackage{rotating,booktabs}
\begin{document}
\begin{frontmatter}

\title{The H$\alpha$ Halo Distribution of 10 Nearby Planetary Nebulae based on the SHASSA Imaging Data}
\author{Chih Hao Hsia}
\address{Institute of Astronomy, National Central University, Chung Li 32054, Taiwan 
         d929001@astro.ncu.edu.tw}
\author{Jin Zeng Li}
\address{National Astronomical Observatories, Chinese Academy of Sciences, Beijing 100012, China
        ljz@bao.ac.cn}
\author{Wing-Huen Ip}
\address{Institute of Astronomy, National Central University, Chung Li 32054, Taiwan}

\begin{abstract}

Here we present initial results of our search for extensive halos
around the planetary nebulae (PNe) in our Galaxy based on imaging data
from the Southern H-Alpha Sky Survey Atlas (SHASSA). A threshold 
surface brightness in H$\alpha$ emission was used to help identify the 
spatially extended features of the PNe. We investigated a sample that included 
10 PNe, the large majority of which were found to be surrounded by extensive outer
halos. The formation of these halos might be closely related to the AGB phase mass loss
and/or the interaction of the stellar outflows with the
interstellar medium. Most of these outer halos are nearly
spherical. Close investigation of some specific objects, such as
He 2-111 and NGC 3242, indicate a kinematic age on the order of
10$^{5}$ yrs. The mass loss history can be traced back to
the late AGB phase of the evolution of the progenitors. Two objects form the
sample (He 2-111, NGC 3242), were found to have outer halos with
fragmented arcs that are apparently the result of their
interaction with the interstellar medium.

\end{abstract}

\begin{keyword}
ISM: general \sep Stars: AGB \sep Planetary Nebulae: Individual 
(NGC 2438 \sep NGC 3242 \sep NGC 7293)
\PACS 97.10.Bt \sep 98.62.Ai \sep 98.62.Mw

\end{keyword}
\end{frontmatter}

\section{Introduction}

Planetary nebulae (PNe) represent one of the most interesting
dynamical phenomena in astrophysics related to the final
stage of low-mass stellar evolution. Kwok et al.
(1978) suggested that the formation of these many-faceted structures, 
characterized by strong emission of H$\alpha$, [NII] and [OIII], is the
direct result of the interaction of a fast wind originating from the nucleus
with a slow wind originatng from the asymptotic giant branch (AGB)
phase of evolution of the progenitor. Magnetic fields, on the
other hand, could also play an important role in shaping 
the various asymmetric structures of the PNe (Blackman et al. 2001).

It is also widely acknowledged that the planetary nebula (PN) is a 
result of an important phase of low-mass stellar evolution that contributes to
the recycling of the interstellar medium (ISM). It is therefore of great 
assistance to our study of the origin of PNe to study the history of
mass loss during the AGB phase and to make a systematic census of
spatially extended halo structures around the PNe. Jewitt
et al. (1986) performed deep exposures of 44 PNe in a H$\alpha$
survey. They found that about two thirds of the observed PNe possessed
extended outer halos. Chu et al. (1987) later showed that
about one-half of their sample of 126 PNe had H$\alpha$ halo structures. 
However, since H$\alpha$ emission from these spatially
extensive halos are intrinsically diffuse and weak, it is highly 
desirable to carry out wide-field surveys with greater sensitivity. 
The release of the Southern H-Alpha Sky Survey
Atlas (SHASSA: \emph{http://amundsen.swarthmore.edu/SHASSA/\/})
(Gaustad et al. 2001) thus offers us a great opportunity to conduct 
a more extensive survey of PNe halos. Here we present the results 
based on this comprehensive data set. The rest of the paper is organized as follow: 
In section 2 we describe the control sample observations. A concise
description of the SHASSA project is presented in section 3. This is
followed by a detailed discussion of some objects in section 4, followed by
a general discussion in section 5. The main results are summarized in section 6.

\section{Observations and data reduction}

We performed deep narrow-band imaging of one of the sources 
in our sample, NGC 2438, to study its halo structures in detail. The
initial mapping of the PN was done by Chu et al. in 1987, but the 
new generation of telescopes makes it easier to resolve the halo structure. 
Our image was compared to the SHASSA survey image. This process serves to 
test the quality and reliability of the SHASSA census of PNe outer halos.

Narrow-band observations were carried out from
2003 to 2005 with the 1 m telescope (LOT) at the Lulin Observatory
in central Taiwan. A FLI IMG 1024S 1024 $\times$ 1024 pixel CCD
and a Princeton Instruments 1340 $\times$ 1300 pixel CCD with a
pixel scale of 0$"$.62 pixel$^{-1}$ were used. The resultant
fields of view were 10$'$ $\times$ 10$'$ and 11$'$
$\times$ 11$'$, respectively. Narrow-band images of NGC 2438
were taken using a H$\alpha$ filter ($\lambda_{c}$ = 6563 \AA,
$\Delta \lambda$ = 30 \AA) with a total exposure time of 16 hr.
Flat fields were obtained at the beginning of the night. The
seeing conditions during the observations ranged between
1$"$.2 and 2$"$.4. A summary of the observations is 
shown in Table 1.

Data reduction included bias and dark current correction,
cosmic ray removal, and flat-fielding, utilizing the standard procedures
in the NOAO IRAF (Ver. 2.12) package. The LOT image was overlaid on the 
SHASSA map. The \emph{ccmap\/} and \emph{ccsetwcs\/} tasks in IRAF 
were used to find the coordinate information.

\section{The SHASSA survey}

The SHASSA data were obtained from a wide-angle H$\alpha$
survey of the southern celestial hemisphere using the robotic El Enano
telescope located at the Cerro Tololo Inter-American Observatory
(\emph{CTIO\/}) in Chile (Gaustad et al. 1999; McCullough et al.
1999). The El Enano telescope has an aperture size of 52 mm, a
focal ratio of f/1.6, and a physical scale of 0$'$.8
pixel$^{-1}$. The field of view is about 13 $^\circ$
$\times$ 13$^\circ$. The narrow-band images were taken through a
H$\alpha$ filter with a bandpass of 3 nm. The emission continuum was 
obtained using the interference filter
with a bandpass of 6 nm on each side of the H$\alpha$.

The exposures were taken in dithered mode to eliminate the ghosts 
of bright stars, the effects of cosmic ray hits, 
and other pixels of abnormal response. 
Background subtraction and median combination were used to process 
the frames. Each co-added image had an angular extension of
roughly 13$^\circ$ $\times$ 13$^\circ$. According to Gaustad et
al. (2001), some residues of bright H$\alpha$ sources could 
still be left behind after continuum subtraction with appropriate
scaling. The sensitivity of the archived SHASSA images were about
0.5 rayleighs (R). One rayleigh is equivalent to $10^{6}$/4$\pi$
photons $cm^{-2}$ $s^{-1}$ $ster^{-1}$, corresponding to an
emission measure of about 2 $cm^{-6}$pc.

Continuum-subtracted images were obtained from the SHASSA archive.
The point spread function (\emph{PSF\/}) could vary from image to
image due to differences in times of observation or color 
of the stellar sources. It is a big challenge to remove
stars from the image without damaging the extended halo structures around the PNe.
We removed residual stellar sources by performing 
linear interpolation of an annulus around the source with a 4
$\sigma$ rejection. This led to striking improvement in the contrast between
the extensive halos and the sky background.
Saturated sources, on the other hand, had to be treated with a 
special procedure. Their bleeding spikes can interfere with the
extended halo structure of the target PNe, and affect the
measurement of their spatial size.

\section{Results and analysis}

We focused on a subset of 10 well-known PNe in the SHASSA
data set. Each had an angular size of greater than 47$"$ and 
a distance of ${\le}$ 1.2 kpc. If extended features surrounding 
each target matched the following criteria, they were taken as 
bona-fide PNe halos: (1) surface brightness
(SB) must be 10 - 10$^{3}$ times fainter than the peak emission of
the main nebula; (2) there was extensive H$\alpha$ emission with an intensity
above 8~$\sigma$ that of the background level; (3) appear as
patches or filaments likely in physical association with the
central PNe; and (4) the spatial extension of the halo is at least
1.6 times larger than the apparent main nebula. It turned out that
the great majority of the sample PNe are surrounded by diffuse
H$\alpha$ halos. To ascertain the extent of halo
structures, and to determine their physical scale, we compared the
SB profiles of the target sources with the average PSF of at least
10 nearby stars with similar brightness. 
To improve the contrast between the bright rim and the faint halo, 
we normalized the SB profile for each PNe to the peak emission of the
rim along particular directions. Table 2 shows the sequence 
number of the target sources, the designation, the coordinates (J 2000.0),
the diameter of the apparent main nebula, previous estimates
of the halo sizes if available, and comments about some specific PNe.

An important factor related to the dynamic evolution of the PNe is
the kinematic age of the halo. The edge of the halo seems to indicate
the location of the last thermal pulse in the AGB phase; the
kinematic age of the halo can be estimated from this event occurred. 
Here the kinematic age (t$_{halo}$) is obtained from the halo size, 
the expansion velocity (V$_{exp}$), and the adopted distance (D), 
all of which are available in the
literature. Due to the fact that historical measurements of the
expansion velocity of the halos are far from complete, we assume an
expansion velocity of 15 km s$^{-1}$, which is the best estimate
for the stellar wind that occurs at the end of the AGB phase of
the evolution of low-mass stars (Habing et al. 1994). Table 3 presents
the above mentioned information and the kinematic ages of the
PNe in our study. A few of them (e.g., He 2-111, NGC 3242 and NGC
7293) have large kinematic ages and are therefore of particular
interest.

\subsection{NGC 2438}

The high quality deep H$\alpha$ images from the LOT observations
(Fig. 1a) provide us with a good opportunity to investigate in detail NGC 2438
and its associated halo structures. Fig. 1a shows the
bright, circular structure (hereafter called the rim) which forms 
the main nebula of NGC 2438. It has a radius of about 25 $"$. It is
noteworthy that the central star does not lie at the geometric
center of the main nebula. There is a second circular structure with an
apparent ridge of H$\alpha$ emission at its edge (hereafter called the
inner halo). It can easily be seen in our deep exposure image. Existing
surveys have already confirmed this nebula's extensive
halo (Balick 1987; Chu et al. 1987; Schwarz et al.
1992; Manchado et al. 1996; Corradi et al. 2000; Corradi et al.
2003). Our deep imaging, however, discloses the existence of a
third faint but distinct structure (hereafter called the outer halo). This
faint halo is also a circular (as delineated by the sharp edge in 
the emission). It is most prominent
between P.A. = -20 $^\circ$ and P.A. = -160 $^\circ$. 
The existence of multiple halo structures hints at an
episodic or periodic mass loss history for NGC 2438.

The \emph{DAOPHOT\/} package of IRAF was used to carry out linear
interpolation of the annulus around each stellar source (with a 3
$\sigma$ rejection) so as to remove all field stars from the image. 
The SB profiles derived from the cleaned image yield an the 
angular radii of the rim, inner and outer halos of
25$"$ (Fig. 1b), 67$"$ and 125$"$ (Fig. 1c),
respectively. As shown in Fig. 1c, the distinct halos have a
relative SB of ${\le}$ 10$^{-1}$ and ${\le}$ 3 $\times$ 10$^{-2}$,
respectively. Baessgen \& Grewing (1989) found a size of
104$"$ for the halo of NGC 2438. Our analysis of the LOT
data, however, resulted in a considerably larger angular scale of
250$"$, which is in agreement with Corradi et al.'s (2000) 
measurement of the order of 230$"$. At a distance of 1 kpc
(Corradi et al. 2000), the angular distances of the inner and
outer halos were 67$"$ and 125$"$, corresponding to 
physical radii of 0.33 pc and 0.6 pc, respectively. Chu \& Jacoby
(1989) gave an expansion velocity of about 15 km s$^{-1}$ from the
halo, which leads to an estimated lifetime of 2.12 $\times$
10$^{4}$ yr and 3.95 $\times$ 10$^{4}$ yr, respectively.

Figure 2 shows the SHASSA contours (starting from 8
$\sigma$ of the background level) overlaid onto the cleaned LOT image. 
The deep exposure image of NGC 2438 confirms the existence of an
extensive outer halo to the PN as well as the validity 
of this search of the SHASSA data for extensive halos 
within a reasonable distance around PNe.

\subsection{He 2-111}

Webster (1978) identified He 2-111 as one of the most
unusual of the PNe, because of the pair of high velocity lobe-like 
structures (with an angular size of about 26 $"$ $\times$ 10$"$), 
on either side of the bright nebular core; PK 315-0.1 (Perek,
1967).

The SB profile of He 2-111 (along the bipolar direction with P.A.
= 144 $^\circ$), and the average PSF of 10 nearby stars with
similar brightness, are shown in Fig. 3a. At an angular resolution
of 0$'$.8 pixel$^{-1}$, the FWHM of the target source is
107$"$. In comparison, the average FWHM of the nearby
field stars with similar brightness is 55$"$. This confirms
the existence of extensive emission in association with the
excitation source of He 2-111. 
To guarantee that only the detection of \emph{bona fide\/} 
halo structures are detected, we must take a threshold 
background level of 8 $\sigma$ in this case which head to 
an estimated radius of the halo of He 2-111 of 340 $"$.

As shown in Figs. 3b and 3c, the extended halo associated with He
2-111 is elongated and the configuration is in alignment with the
bipolar distribution of the nebula. The SHASSA image of the nebula
indeed indicates enhanced H$\alpha$ emission at the tips of the
bipolar lobes. This suggests interaction with the
surrounding ISM. If a distance of 1.2 kpc (Phillips 2002) is
adopted, the physical scale of the halo would be about 3.96 pc.
When an expansion velocity of 15 km s$^{-1}$ (Habing et al. 1994)
is assumed, the kinematic age of He 2-111 is estimated to be 1.3
$\times$ 10$^{5}$ yrs.

\subsection{NGC 3242}

It was Deeming (1966) who first identified the faint arc-like filament, 
about 10 $'$ long, southwest of NGC 3242. Kaftan-Kassim (1966)
argued that this arc structure is likely in physical association
with the PN based on 1400 MHz radio mapping of NGC 3242. 
Later, images from the Palomar Sky Survey (POSS)
(see Figure 4c) suggested that the arc-like filament was merely a
portion of an elliptical halo (with a spatial extension of about 18
$'$ $\times$ 24 $'$) (Bond 1981).

Figure 4a shows a comparison of the SB profile of the excitation
source of NGC 3242 along P.A. = 173 $^\circ$ and the average PSF
of 10 nearby field stars with similar brightness. The FWHM of NGC
3242 was 106 $"$ and that of the field stars 70$"$. 
This confirmed the extensive nature of NGC 3242. Our study of 
the SHASSA survey, however, shows the halo of NGC 3242 to be 
much more extensive $\sim$ 36.7 $'$ (see Figs. 4b
and 4c), that is about 4.48 pc in scale for a distance of 420 pc (Hajian
et al. 1995). Our analysis of the SHASSA data showed NGC 3242 
to have a larger halo than that found by Corradi et al. (2003), who 
found a halo size of $\sim$ 33 $'$. The implication is that 
isotropic mass outflow from the progenitor has rammed into clumpy
ISM, residing predominantly in the south-west of the PN, which 
produced the swept-up appearance of this object. 
This outflow was less disturbed in other directions. The
fine structures displayed are likely a result of some kind of
instability in the dissipation of the halo. The arc
structures indicate a displacement in radial velocity of 16 km
s$^{-1}$, as compared to the systematic velocity of NGC 3242
(Meaburn et al. 2000). The estimated kinematic age of 
NGC 3242 is thus about 1.37 $\times$ 10$^{5}$ yrs, which 
places its origin firmly into the AGB phase of the evolution of
the progenitor.

\subsection{Helix Nebula (NGC 7293)}

The peculiar optical appearance at the Helix Nebula 
(NGC 7293; PK 36-57.1) makes it among the most well
known of the bipolar PNe.
Its close proximity, a distance of 213 pc away (Harris et al. 1997),
and its large angular size $\sim$ 10$'$ (Pottasch
1984), make it most suitable case for a study of the spatial
distribution and kinematics of the multiple envelopes of evolved
PNe.

Figure 5a. shows a plot of the SB profile of the Helix Nebula 
along P.A. = 122 $^\circ$ and the average
PSF of 10 surrounding stars of similar brightness. 
When the angular resolution is 0$'$.8 pixel$^{-1}$,
the FWHM of NGC 7293 is 595$"$, nearly an
order of magnitude larger than the value of 71$"$ obtained for the field
stars of similar brightness. Chu et al. (1987) gave an
estimated optical diameter of $\sim$ 24.3$'$ for the Helix Nebula. 
Young et al. (1999), on the other hand, gave an angular
size of $\sim$ 16.7$'$ for the molecular envelope of the
nebula. Based on the SHASSA survey images, however, we found
the halo size to be as much as 44.3$'$, a factor of two
larger than those obtained previously (see Figs. 5b and 5c).
At a distance of 213 pc (Harris et al. 1997), the
physical scale of the halo is 2.74 pc. The kinematic age of 
the Helix Nebula is estimated to be $\sim$ 8.95 $\times$ 10$^{4}$ yrs,
if the expansion velocity of the halo is 15 km s$^{-1}$ (Habing et
al. 1994).

\section{Discussion}

\subsection{Surface brightness}

Figures. 6a-11a show the SB profiles for 6 other well-observed PNe 
(IC 4406, NGC 2818, NGC 2899, NGC 5189, NGC 6072, NGC 6302) along 
different P.A.. Figs 6b-11b show the contours of
their corresponding SHASSA H$\alpha$ images overlaid on the DSS II
red plate images. An examination of the histograms of the SB 
distribution of the halos relative to the peak emission of each PN 
shows that in the majority of the extensive halos detected, the relative 
SB is less than 10$^{-1}$. One exception is He 2-111, which has a relative SB
of $\sim$ 2 $\times$ 10$^{-1}$. According to Villaver (2001), the
enhanced emission at the tip of the bipolar lobes of He 2-111 can be 
attributed to their interaction with the ISM.

\subsection{Properties of the AGB halos}

In our sample the large majority of the extensive halos surrounding 
the PNe are well-defined in shape and on a physical scale of up to a
few parsecs. Their largely smooth and isotropic distribution
indicates that they most likely experienced a similar mass loss 
history during the AGB phase of evolution. However, some PNe, 
such as NGC 3242 (Fig. 4) and He 2-111 (Fig. 3), have bow-shocked and/or 
filamentary halo structures. High resolution HST imaging reveals
that the central regions of the proto-planetary nebulae (PPNe) show
bipolar symmetry (Kwok et al. 2000). This hints at a change in
the mode of the mass loss of these objects, either at the end of their AGB
evolution or in the early PPNe phase. The shaping process could be 
a result from external torque caused by the emergence of
magnetic fields, or perhaps due to a close binary origin of the
central source producing a strong outgoing wind, as suggested by the
studies of Hb 12 (Hsia et al. 2006), MyCn 18 (Bryce et al. 1997),
and Red Rectangle Nebula (Van Winckel et al. 1995).

Table 2 shows that our PNe halo sizes are much larger than 
those previously estimated. The presence of extended halos 
indicates that the progenitors of the PNe were subjected 
to extensive episodes of mass loss during their AGB 
evolutionary phase.

Assuming a spherical geometry, we could compute the kinematic ages 
of the ionized halos from the apparent size, distance, and
expansion velocity (Guerrero et al. 1998) available in the
literature. The derived kinematic age of the ionized halos
agrees well with the timescale of the AGB phase of the evolution of
the progenitors.

\subsection{Interaction of PNe with ISM}

The interaction of PNe with the ISM was first described by
Gurzadyan (1969) and later discussed in more detail by Smith
(1976). In the past three decades, there have been extensive
studies on the process of mass loss by which PNe supply material
to the ISM (e.g., Balick 1987; Frank \& Mellema 1994; Mellema
1993). Dgani \& Soker (1998) suggested that the morphology of
interacting PNe may be a result of Rayleigh-Taylor (RT) instability, 
resulting in fragmentation at the halos of the PNe. The magnetic field 
could play an important role in this interaction 
(Tweedy et al. 1995; Soker \& Dgani 1997; Zucker \& Soker 1997), 
and should not be ignored in the analysis (Soker \& Dgani 1997).

According to Tweedy \& Kwitter (1996) and Dgani \& Soker (1998), 
the interaction of the PNe with the ISM has the following distinct
characteristics: (1) the outer regions of the
nebulae are asymmetric; (2) there is flux enhancement in the outer
regions of the nebula accompanied by a drop in the ionization
level; (3) there is fragmentation of the halo or the presence of
filamentary arcs. Sources were selected from the sample such as 
He 2-111, and NGC 3242 as probable candidates showing the interaction 
of the PNe. Some of the impressive morphological features of these 
objects are revealed for the first time. For example in Figure 3 we see 
He 2-111 shows a pair of bipolar structures and, in Figure 4 (NGC 3242) 
we see a prominent arc-like structure; 
both are typical features of interacting PNe.

\section{Summary}

Our investigation of a sample of 10 known PNe, taken primarily 
from SHASSA data, has led to the following results:

\begin{enumerate}
  \item The large majority of objects in our sample were found 
to be surrounded by extensive outer halos that likely originated from AGB phase
mass loss. Some of the sample sources, such as He 2-111 and NGC
3242, had outer halos with prominent arcs and/or filamentary
structures, suggesting apparent interaction with the ISM.
  \item Most of the outer halos were isotropic. Close investigation
of some specific objects (He 2-111, and NGC 3242) indicate that 
they had a kinematic age on the order of 10$^{5}$ yrs. This allows us 
to trace the mass loss history back to the late AGB 
phase of the evolution of the progenitors.

\end{enumerate}

{\flushleft \bf Acknowledgments~}

This work is based on an exploration of the Southern H-Alpha Sky
Survey Atlas (SHASSA), which was financially supported by the National Science
Foundation, USA. We are grateful to Jing-Yao Hu at the \emph{National
Astronomical Observatory of the Chinese Academy of Sciences\/} for 
his useful comments and discussions. This project is partially
supported by the "\emph{National Science Council of Taiwan\/} under
NSC 94-2752-M-008-001-PAE, and NSC 94-2112-M-008-002", and by the
"Aim for the Top University Program" of the Ministry of Education,
Taiwan. Finally, we acknowledge funding from \emph{the National
Natural Science Foundation of China\/} through grant 10503006.

\clearpage

\begin{table}
\caption{Journal of H$\alpha$ Observations of NGC 2438}
\label{LOTobservation}
\begin{tabular}{ccccc}
\hline\hline
Observation Date & Exposure Time (s) & Number of Exposures & Seeing ($"$) & Duration (hr) \\
\hline
2003 Mar. 4 & 300 & 21 & 1.4 & 1.75 \\
2003 Mar. 8 & 300 & 34 & 1.2 $\sim$ 1.8 & 2.8 \\
2003 Mar. 12 & 300 & 26 & 1.2 $\sim$ 1.7 & 2.17 \\
2003 Mar. 13 & 300 & 13 & 1.5 & 1.08 \\
2005 Jan. 11 & 600 & 21 & 1.4 $\sim$ 1.9 & 3.5 \\
2005 Jan. 12 & 600 & 18 & 1.5 $\sim$ 2 & 3  \\
2005 Jan. 14 & 600 & 9 & 1.5 $\sim$ 2.4 & 1.5 \\
\hline
\end{tabular}
\end{table}
\clearpage

\begin{sidewaystable}
\caption{Extensive halos associated with the 10 southern PNe} \label{HaloPNe}
\begin{tabular}{cccccccc}
\hline\hline 
Number & Object & RA (J 2000.0) & DEC (J 2000.0) & Main Nebula ($"$) & Previous Halo Size ($"$) &
Halo Size ($"$) & Note \\
\hline
1 & NGC 7293 & 22 29 38.6 & -20 50 13.6 & 1460$^{a}$ & 1350$^{e}$ & 2660 &  \\
2 & NGC 2438 & 07 41 51.4 & -14 43 54.9 & 152$^{c}$ & 104$^{b}$, 226$^{e}$ & 250 & 2 \\
3 & NGC 3242 & 10 24 46.1 & -18 38 32.6 & 47$^{a}$ & 150$^{b}$ , $>$194 $^{e}$ &
2200 & 1 \\
4 & NGC 2818 & 09 16 01.7 & -36 37 38.8 & 123 $\times$ 55$^{d}$ & -- & 288 &  \\
5 & NGC 2899 & 09 27 03.1 & -56 06 21.2 & 140 $\times$ 69$^{d}$ & -- & 238 &  \\
6 & NGC 5189 & 13 33 33.0 & -65 58 26.7 & 185 $\times$ 152$^{d}$ & -- & 302 &  \\
7 & He 2-111 & 14 33 18.4 & -60 49 35.0 & $>$143 $\times$ 76$^{d}$ & -- & 680 & 1 \\
8 & IC 4406 & 14 22 26.3 & -44 09 04.4 & 103$^{c}$ & -- & 282 & \\
9 & NGC 6072 & 16 12 58.1 & -36 13 46.1 & $>$145$^{c}$ & -- & 246 & \\
10 & NGC 6302 & 17 13 44.2 & -37 06 15.9 & $>$155$^{c}$ & -- & 278 & \\
\hline
\end{tabular}
\\
\begin{list}{}{}
\item[] (1) Known interaction with ISM\ \item[] (2) 2nd halo\
\item[] \item[] (a) Chu et al. 1987\ \item[] (b) Baessgen \&
Grewing 1989\ \item[] (c) Schwarz et al. 1992\ \item[] (d)
G$\acute{o}$rny et al. 1999\ \item[] (e) Corradi et al. 2003\
\end{list}
\end{sidewaystable}
\clearpage

\begin{sidewaystable}
\caption{Parameters of the halos of 10 PNe in the Milkway Galaxy}
\label{10PNe}
\begin{tabular}{cccccccc}
\hline\hline Number & Object & l , b & D$^{a}$ (kpc) & Halo Size
($'$) & V$_{exp}$$^{b}$ (km s$^{-1}$) & t$_{halo}$ ($\times$
10$^{-4}$ yrs) & Note \\
\hline
1 & NGC 7293 & 036.1-57.1 & 0.213$^{c}$ & 44.34 & 15 & 8.95 &   \\
2 & NGC 2438 & 231.8+04.1 & 1$^{d}$ & 2.23 & 15$^{e}$ & 2.12 & inner halo \\
  &          &            &                    & 4.17 & 15$^{e}$ & 3.95 & outer halo \\
3 & NGC 3242 & 261.0+32.0 & 0.42$^{f}$ & 36.67 & 16$^{g}$ & 13.7 &  \\
4 & NGC 2818 & 261.9+08.5 & 1.14 & 4.8 & 15 & 5.19 &   \\
5 & NGC 2899 & 277.1-03.8 & 0.69 & 3.96 & 15 & 2.6 &   \\
6 & NGC 5189 & 307.2-03.4 & 0.33 & 5.04 & 15 & 1.58 &   \\
7 & He 2-111 & 315.0-00.3 & 1.2 & 11.34 & 15 & 12.9 &  \\
8 & IC 4406 & 319.6+15.7 & 0.79 & 4.7 & 15 & 3.52 &  \\
9 & NGC 6072 & 342.2+10.8 & 0.57 & 4.1 & 15 & 2.22 &  \\
10 & NGC 6302 & 349.5+01.1 & 0.21 & 4.64 & 15 & 0.93 & \\
\hline
\end{tabular}
\begin{list}{}{}
\item[] \item[] (a) Phillips, J. P. 2002\ \item[] (b) Habing et
al. 1994\ \item[] (c) Harris et al. 1997\ \item[] (d) Corradi et
al. 2000\ \item[] (e) Chu \& Jacoby 1989\ \item[] (f) Hajian et
al. 1995\ \item[] (g) Meaburn et al. 2000\
\end{list}
\end{sidewaystable}
\clearpage

 \begin{figure}
   \centering
   \includegraphics[width=14cm]{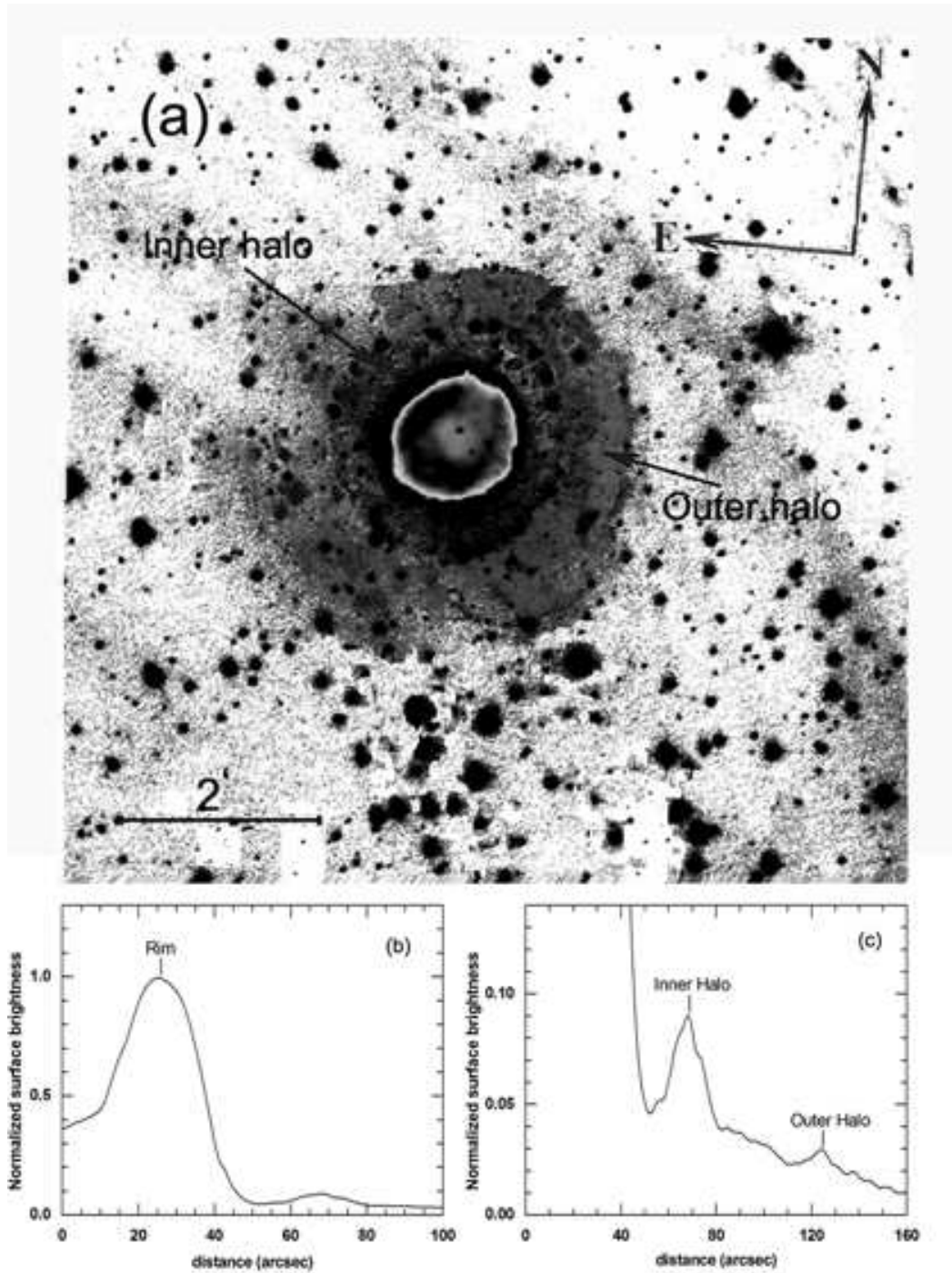}
      \caption{(a) The narrowband image of
                  NGC 2438 taken with LOT. The image utilizes a
                  logarithmic gray scale. The circular inner halo and the outer halo
                  are clearly shown. (b) Normalized SB profile of
                  the rim calculated between position angle (P.A.) of -30 $^\circ$ and  -120
                  $^\circ$. (c) Normalized SB profile of the inner halo and outer
                  halo measured between P.A. = -30 $^\circ$ and P.A. = -120 $^\circ$.
              }
         \label{Figure 1}
   \end{figure}

 \begin{figure}
   \centering
   \includegraphics[width=16cm]{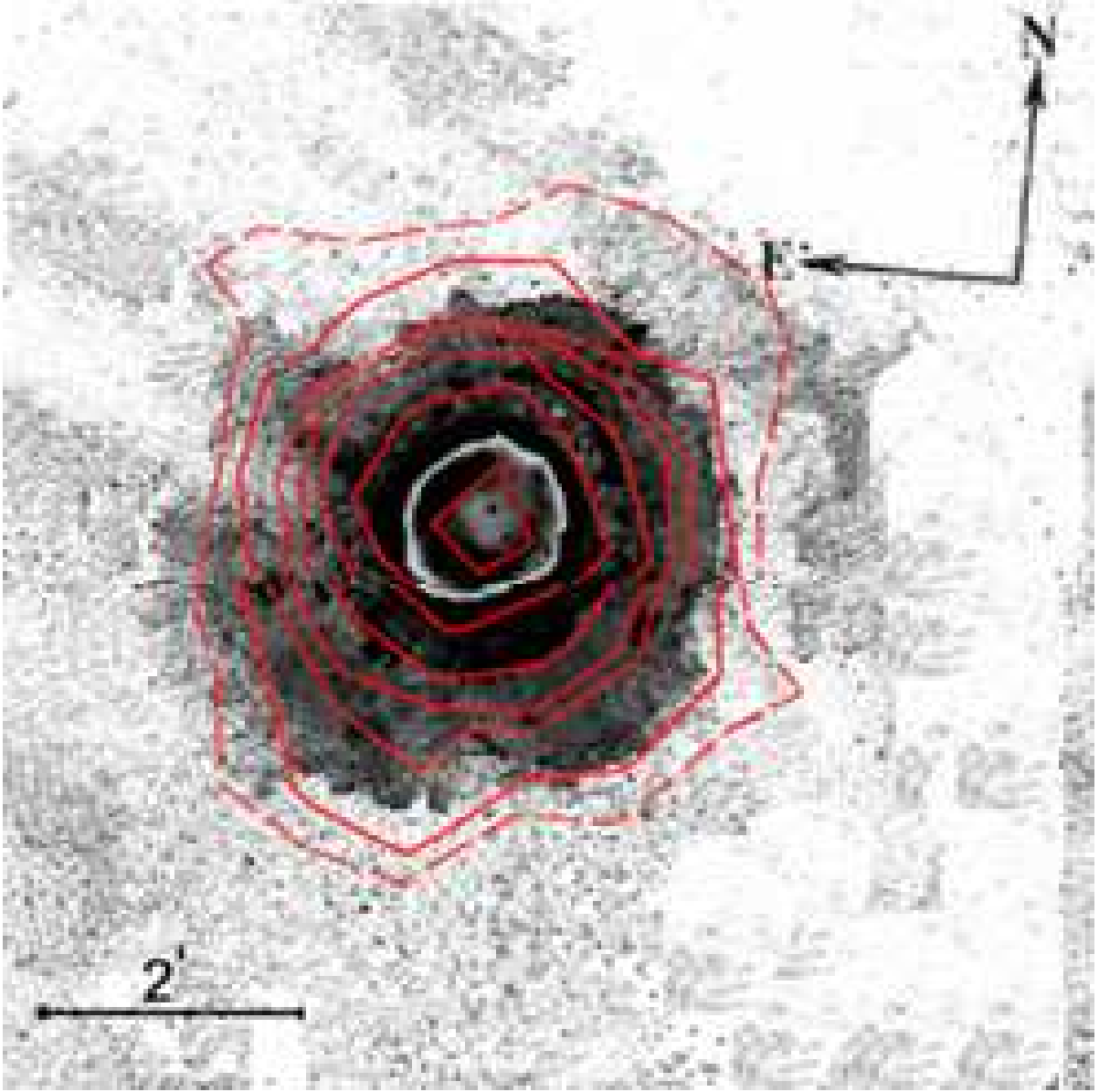}
      \caption{Distribution of the H$\alpha$ halo of NGC
               2438. Contour plot of the SHASSA H$\alpha$ image was overlaid
               on the cleaned LOT image of the PN. The lowest level dotted contour corresponds
               to 8.09 $\times$ 10$^{-16}$ ergs $cm^{-2}$ $s^{-1}$ $arcsec^{-2}$,
               3 $\sigma$ above the background. The solid contours are 2.15
               $\times$ 10$^{-15}$ (8 $\sigma$ above the background), 4.31
               $\times$ 10$^{-15}$, 8.62 $\times$ 10$^{-15}$, 1.72 $\times$
               10$^{-14}$, 3.45 $\times$ 10$^{-14}$, and 6.89 $\times$ 10$^{-14}$
               ergs $cm^{-2}$ $s^{-1}$ $arcsec^{-2}$, respectively.
               }
         \label{Figure 2}
   \end{figure}

\begin{figure}
   \centering
   \includegraphics[width=11.5cm]{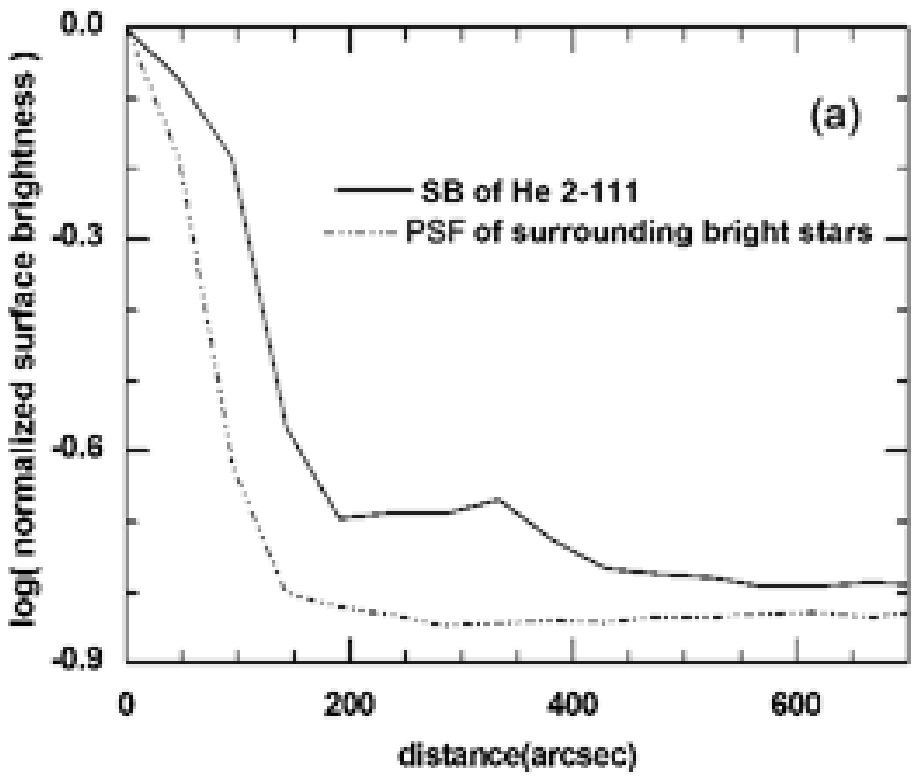}
   \includegraphics[width=6.9cm]{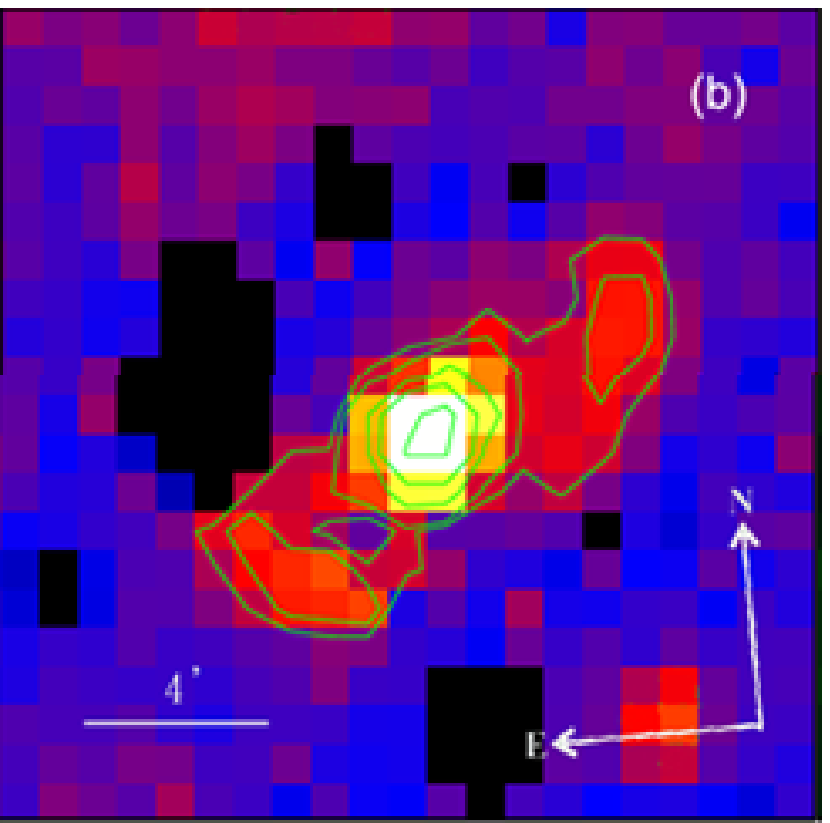}
   \includegraphics[width=6.8cm]{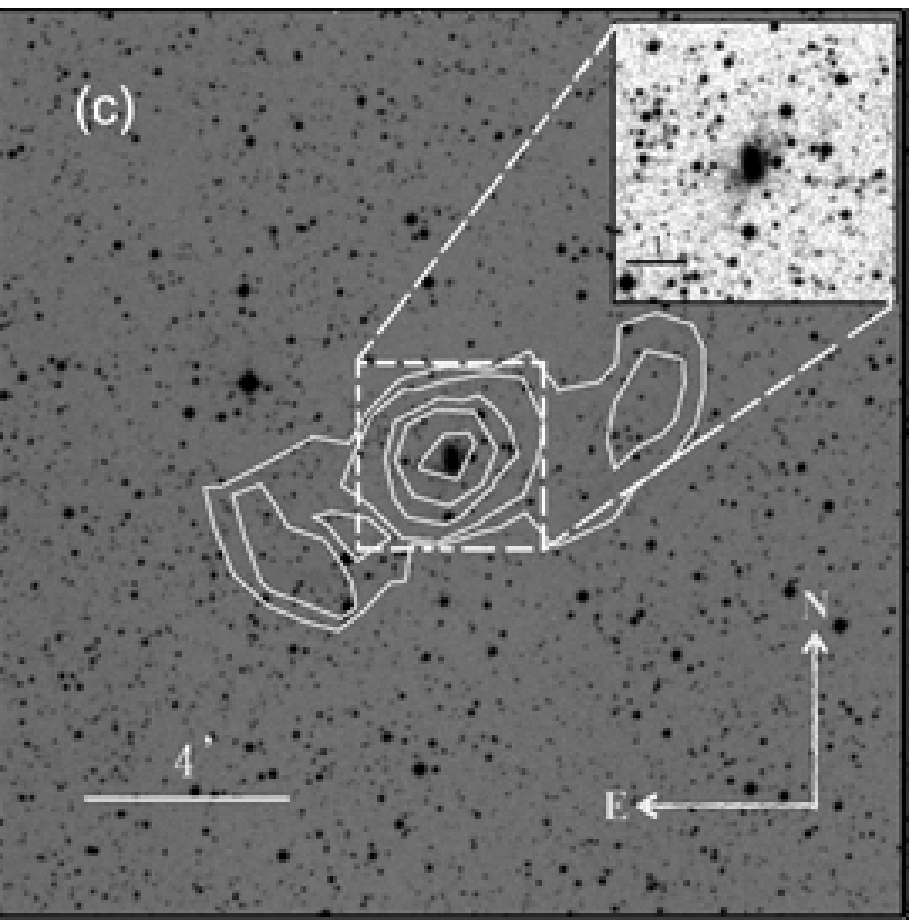}
      \caption{(a) Normalized SB profile of
                   the halo of He 2-111 along P.A. = 144 $^\circ$ (solid line) and
                   the average PSF of nearby stars with similar brightness (dotted
                   line). (b) SHASSA H$\alpha$ image of He 2-111 with surface
                   brightness contours. (c) Contours of H$\alpha$ emission of the PN
                   overlaid on the DSS II plate image with a spatial extension of
                   19.1$'$ $\times$ 19.1$'$. The lowest contour level
                   corresponds to 2.92 $\times$ 10$^{-15}$ ergs $cm^{-2}$ $s^{-1}$
                   $arcsec^{-2}$, 8 $\sigma$ above the background. The other contours
                   are 4.51 $\times$ 10$^{-15}$,
                   1.08 $\times$ 10$^{-14}$, 2.6 $\times$ 10$^{-14}$, and 6.24
                   $\times$ 10$^{-14}$ ergs $cm^{-2}$ $s^{-1}$ $arcsec^{-2}$,
                   respectively. The inset shows the enlarged image of the central
                   region of the nebula.
              }
         \label{Figure 3}
   \end{figure}

\begin{figure}
   \centering
   \includegraphics[width=11.5cm]{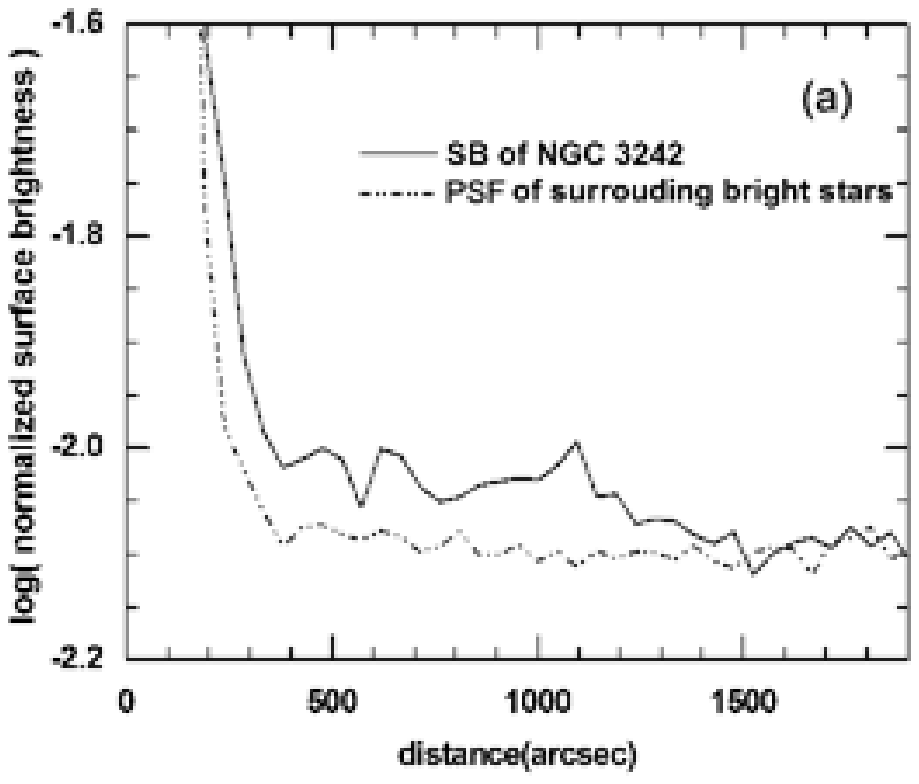}
   \includegraphics[width=6.9cm]{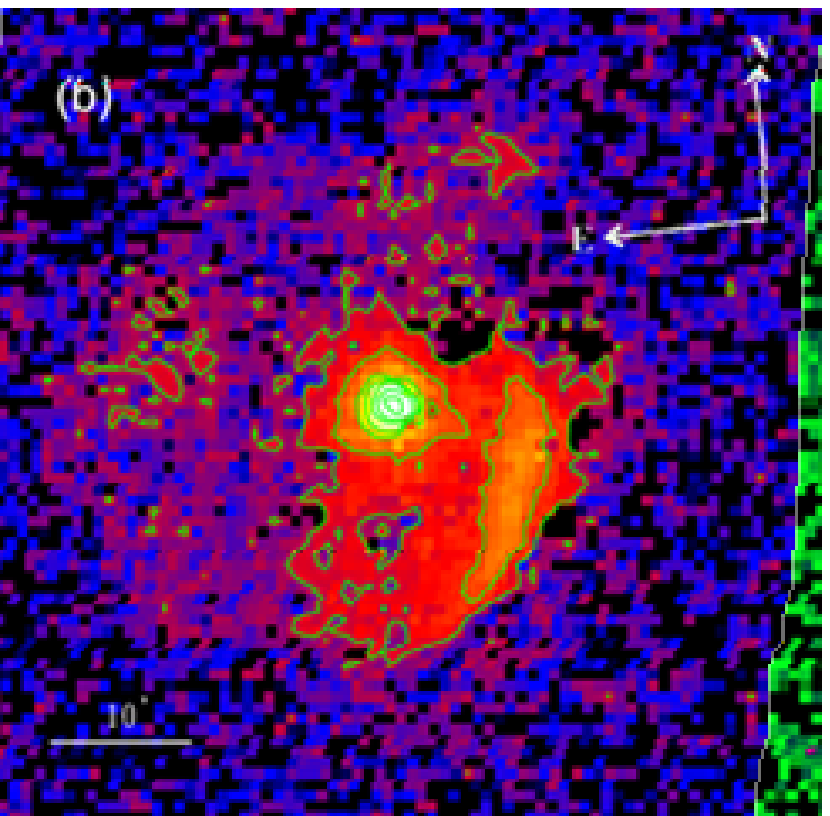}
   \includegraphics[width=6.8cm]{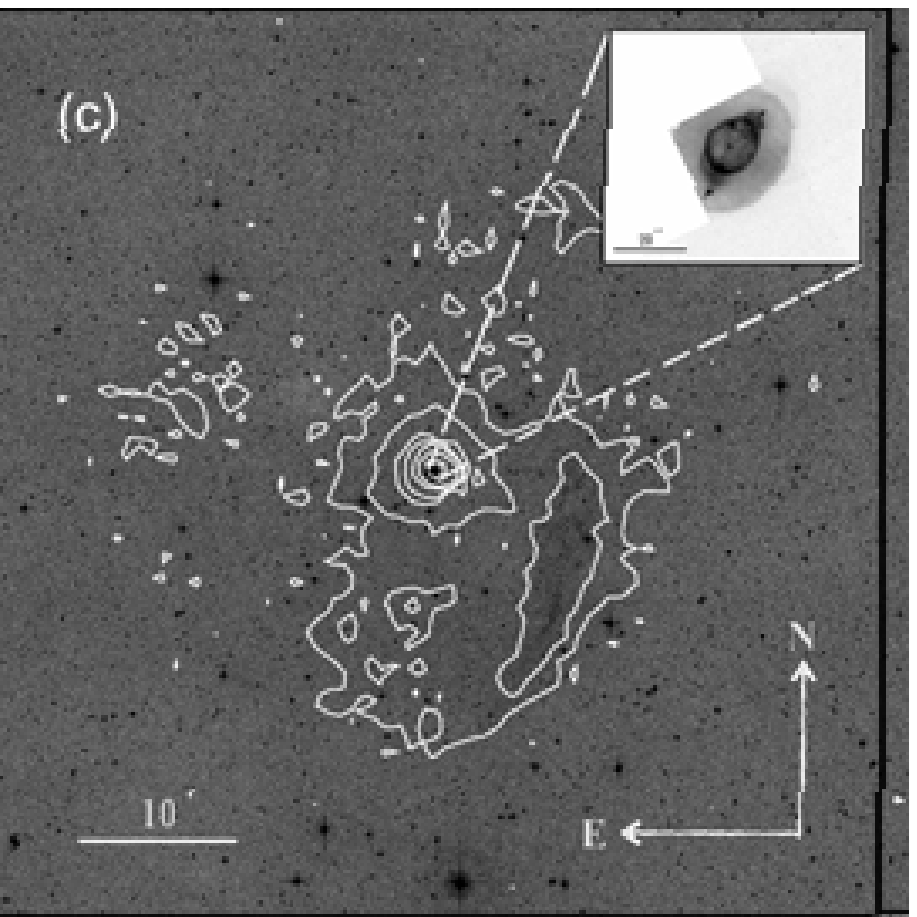}
      \caption{(a) Normalized SB profile of
of NGC 3242  along P.A. = 173 $^\circ$ (solid line) and the
average PSF of surrounding stars with similar brightness (dotted
line). (b) the SHASSA H$\alpha$ image with contours of the surface
brightness distribution overlaid. (c) Contours of the SHASSA H$\alpha$ 
PN image superimposed onto the DSS II plate image, with an
extension of 59.6$'$ $\times$ 59.6$'$ centered on NGC
3242. The lowest contour level corresponds to 6.3 $\times$
10$^{-16}$ ergs $cm^{-2}$ $s^{-1}$ $arcsec^{-2}$, 8 $\sigma$ above
the background. The other contours are 1.98 $\times$ 10$^{-15}$,
6.26 $\times$ 10$^{-15}$, 1.98 $\times$ 10$^{-14}$, 6.25 $\times$
10$^{-14}$, 1.98 $\times$ 10$^{-13}$, and 6.24 $\times$ 10$^{-13}$
ergs $cm^{-2}$ $s^{-1}$ $arcsec^{-2}$, respectively. The inset
shows an enlarged [NII] HST image of the main nebula. }
         \label{Figure 4}
   \end{figure}

\begin{figure}
   \centering
   \includegraphics[width=11.5cm]{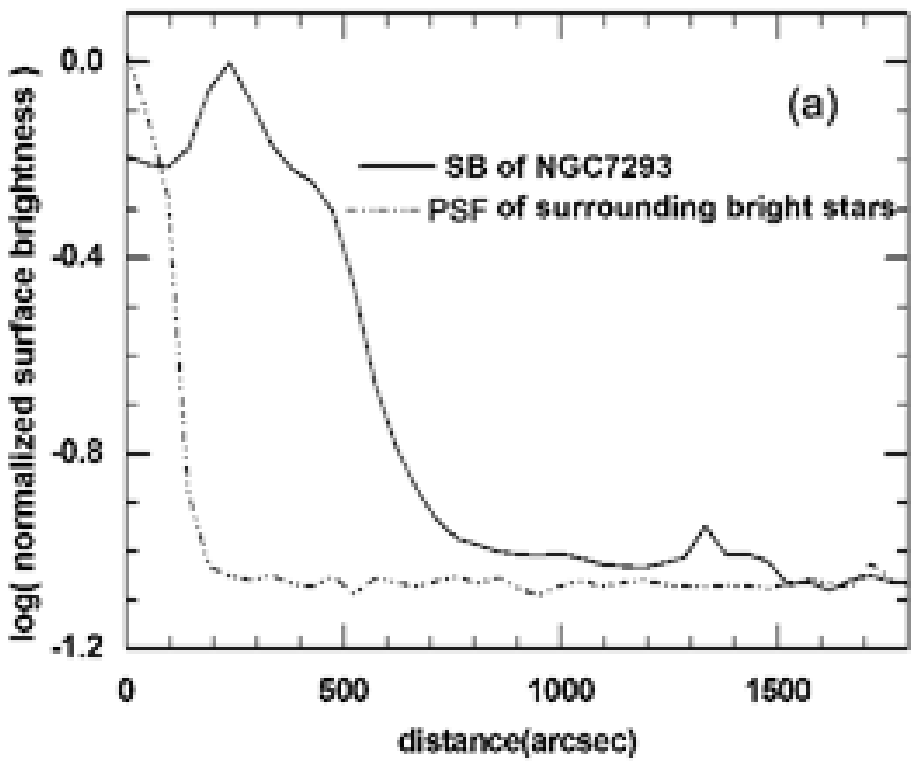}
   \includegraphics[width=6.9cm]{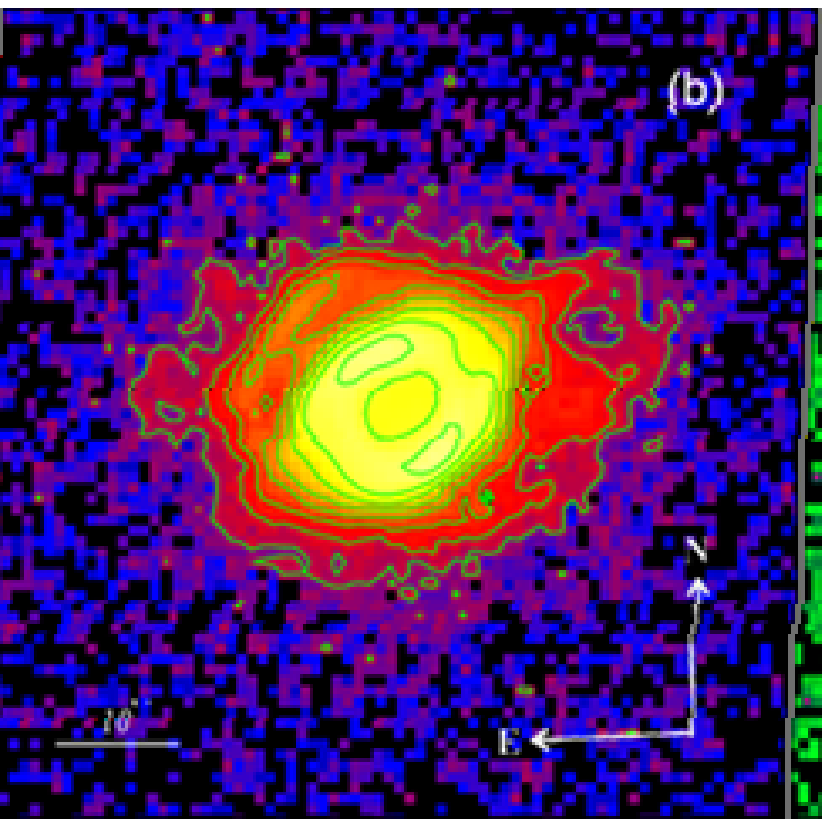}
   \includegraphics[width=6.8cm]{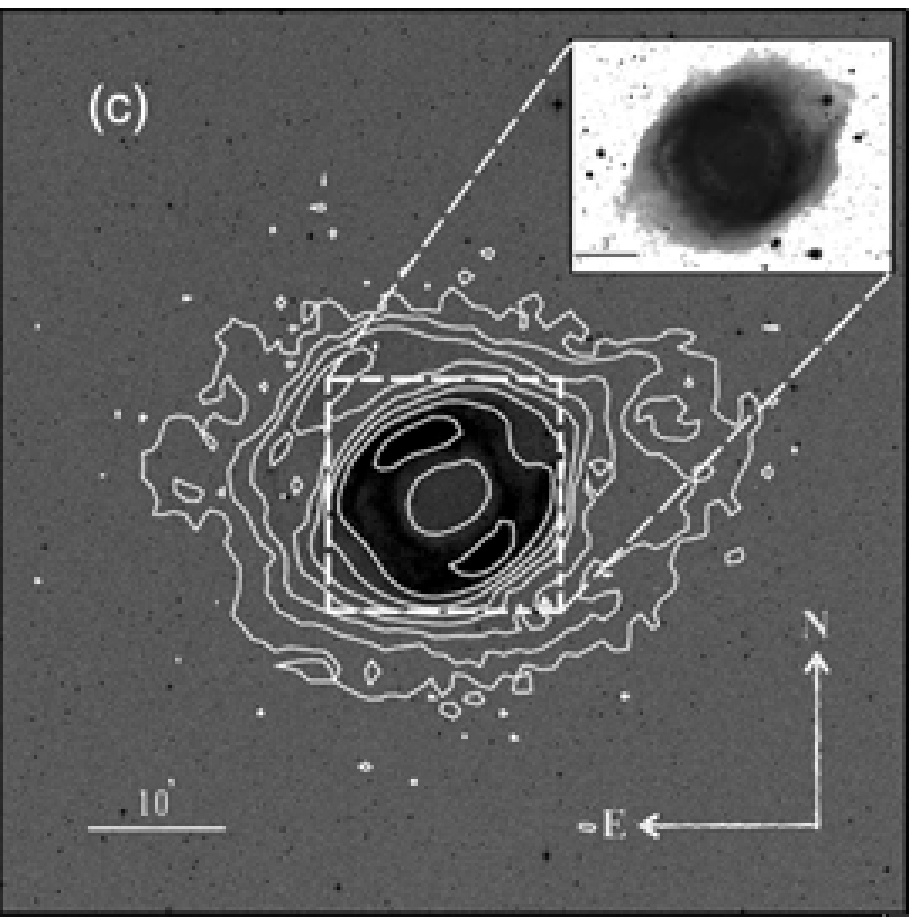}
      \caption{(a) The normalized SB profile
of the halo of NGC 7293  along P.A. = 122 $^\circ$ (solid line)
and the average PSF of surrounding field stars with similar
brightness (dotted line). (b) The SHASSA H$\alpha$ image of NGC
7293 with superimposed contours of the surface brightness
distribution. (c) Contour plot of the SHASSA H$\alpha$ imaging
overlaid onto a 59.6$'$ $\times$ 59.6$'$ DSS II red plate
image. The lowest level contour corresponds to 6.1 $\times$
10$^{-16}$ ergs $cm^{-2}$ $s^{-1}$ $arcsec^{-2}$, 8 $\sigma$ above
the background. The other contours are 1.32 $\times$ 10$^{-15}$,
2.77 $\times$ 10$^{-15}$, 5.8 $\times$ 10$^{-15}$, 1.22 $\times$
10$^{-14}$, 2.56 $\times$ 10$^{-14}$, 5.38 $\times$ 10$^{-14}$,
and 1.13 $\times$ 10$^{-13}$ ergs $cm^{-2}$ $s^{-1}$
$arcsec^{-2}$, respectively. The inset shows an enlarged CFHT
image of the main nebula. }
         \label{Figure 5}
   \end{figure}

   \begin{figure}
   \centering
   \includegraphics[width=10cm]{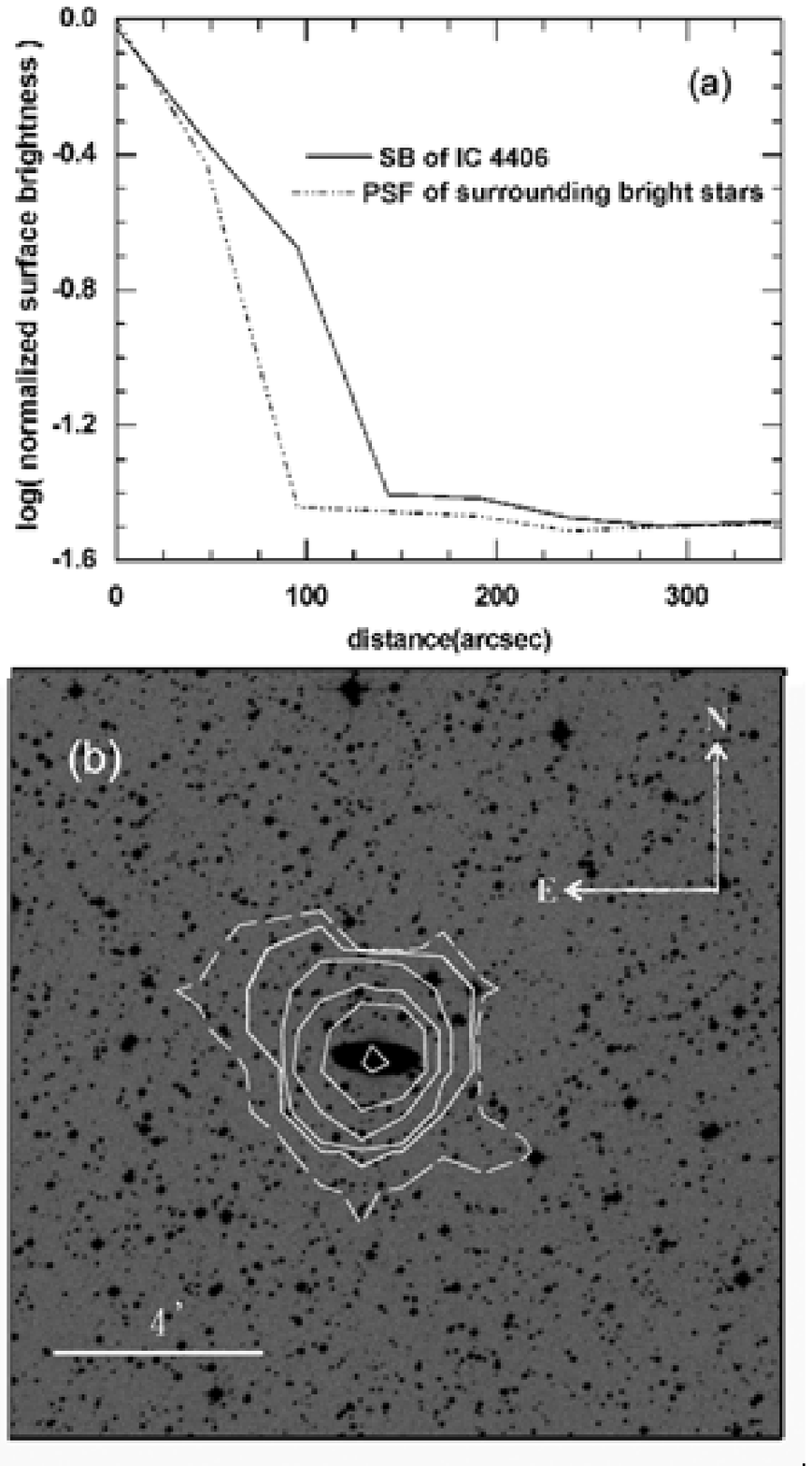}
      \caption{(a) Normalized SB profile of the halo
for IC 4406  along P.A. = 90 $^\circ$ (solid line) and the average
PSF of nearby stars with similar brightness (dotted line). (b)
Contours of SHASSA H$\alpha$ imaging of the PN overlaid on the DSS
II red plate image with a spatial extension of 18.6$'$ $\times$
18.6$'$. The lowest level dotted contour corresponds to 7.78
$\times$ 10$^{-16}$ ergs $cm^{-2}$ $s^{-1}$ $arcsec^{-2}$, 3
$\sigma$ above the background. The other solid contours are 2.07
$\times$ 10$^{-15}$ (8 $\sigma$ above the background), 6.56
$\times$ 10$^{-15}$, 2.07 $\times$ 10$^{-14}$, 6.55 $\times$
10$^{-14}$, and 2.07 $\times$ 10$^{-13}$ ergs $cm^{-2}$ $s^{-1}$
$arcsec^{-2}$, respectively.}
         \label{Figure 6}
   \end{figure}

\begin{figure}
   \centering
   \includegraphics[width=10cm]{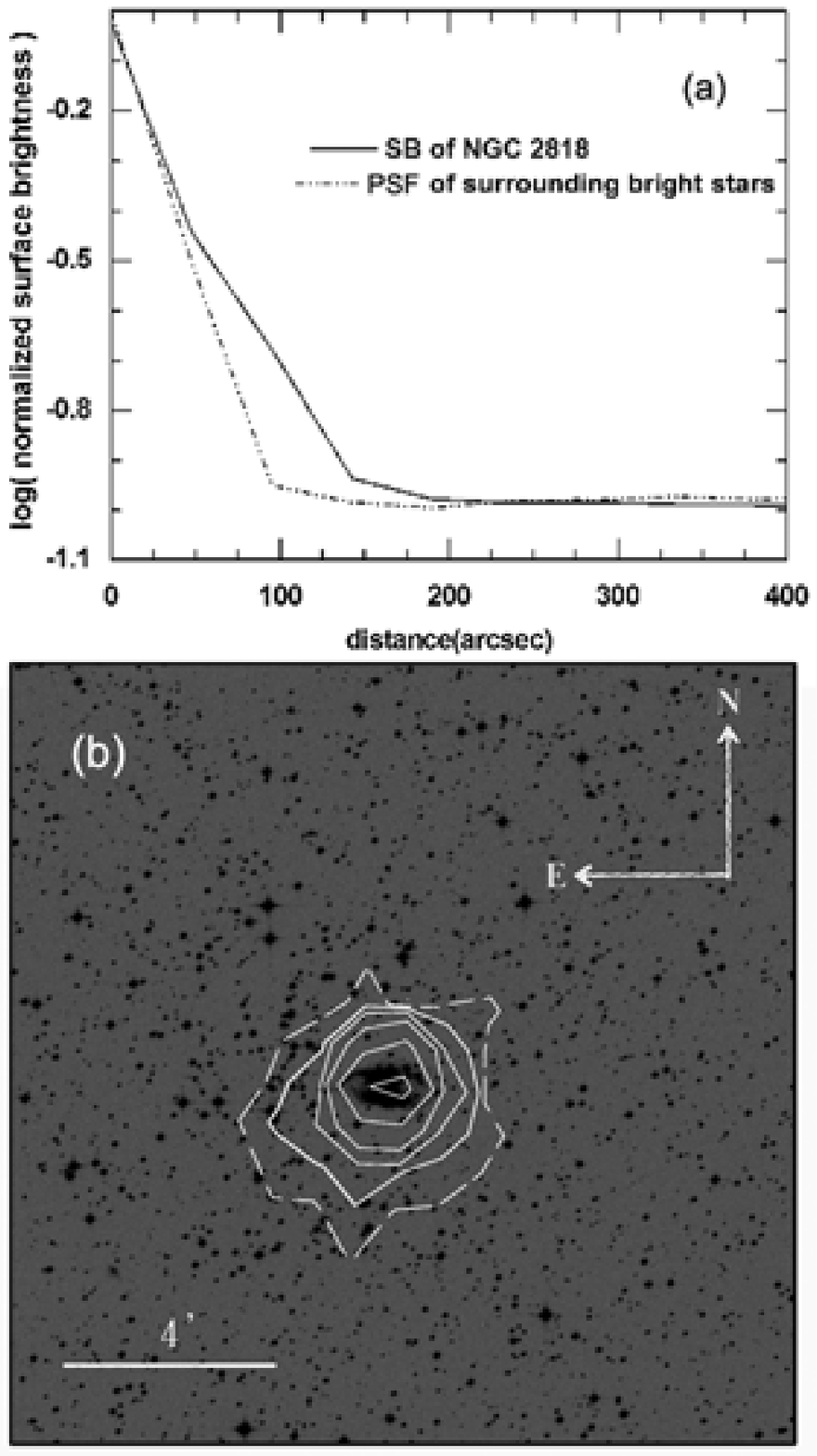}
      \caption{(a) The normalized SB profile of the
halo of NGC 2818 along P.A. = 90 $^\circ$ (solid line) and the
average  PSF of nearby stellar sources with similar brightness
(dotted line).  (b) Contour plot of the SHASSA H$\alpha$ imaging
overlaid on a 18.6$'$ $\times$ 18.6$'$ DSS II red plate
image. The lowest level dotted contour corresponds to 1.1 $\times$
10$^{-15}$ ergs $cm^{-2}$ $s^{-1}$ $arcsec^{-2}$, 3 $\sigma$ above
the background. The other solid contours are 2.93 $\times$
10$^{-15}$ (8 $\sigma$ above the background), 6.44 $\times$
10$^{-15}$, 1.42 $\times$ 10$^{-14}$, 3.12 $\times$ 10$^{-14}$,
and 6.86 $\times$ 10$^{-14}$ ergs $cm^{-2}$ $s^{-1}$
$arcsec^{-2}$, respectively.}
         \label{Figure 7}
   \end{figure}

\begin{figure}
   \centering
   \includegraphics[width=10cm]{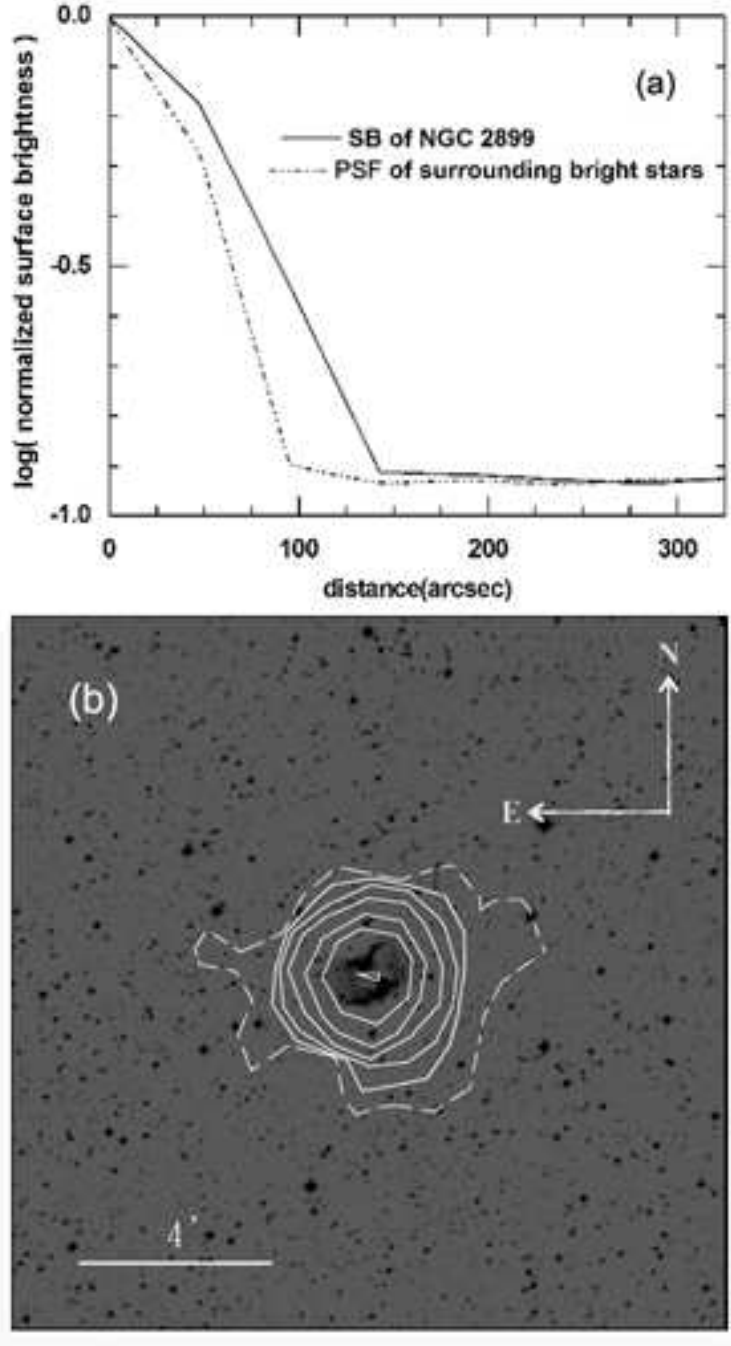}
      \caption{(a) The normalized SB profile of the
halo of NGC 2899  along P.A. = 90 $^\circ$ (solid line) and the
average PSF of nearby stars with similar brightness (dotted line).
(b) Contour plot of the SHASSA imaging overlaid on a 18.6$'$
$\times$ 18.6$'$ DSS II red plate image. The lowest level
dotted contour corresponds to 1.45 $\times$ 10$^{-15}$ ergs
$cm^{-2}$ $s^{-1}$ $arcsec^{-2}$, 3 $\sigma$ above the background.
The other solid contours are 3.88 $\times$ 10$^{-15}$ (8 $\sigma$
above the background), 7.75 $\times$ 10$^{-15}$, 1.55 $\times$
10$^{-14}$, 3.1 $\times$ 10$^{-14}$, 6.2 $\times$ 10$^{-14}$, and
1.24 $\times$ 10$^{-13}$ ergs $cm^{-2}$ $s^{-1}$ $arcsec^{-2}$,
respectively.}
         \label{Figure 8}
   \end{figure}

\begin{figure}
   \centering
   \includegraphics[width=10cm]{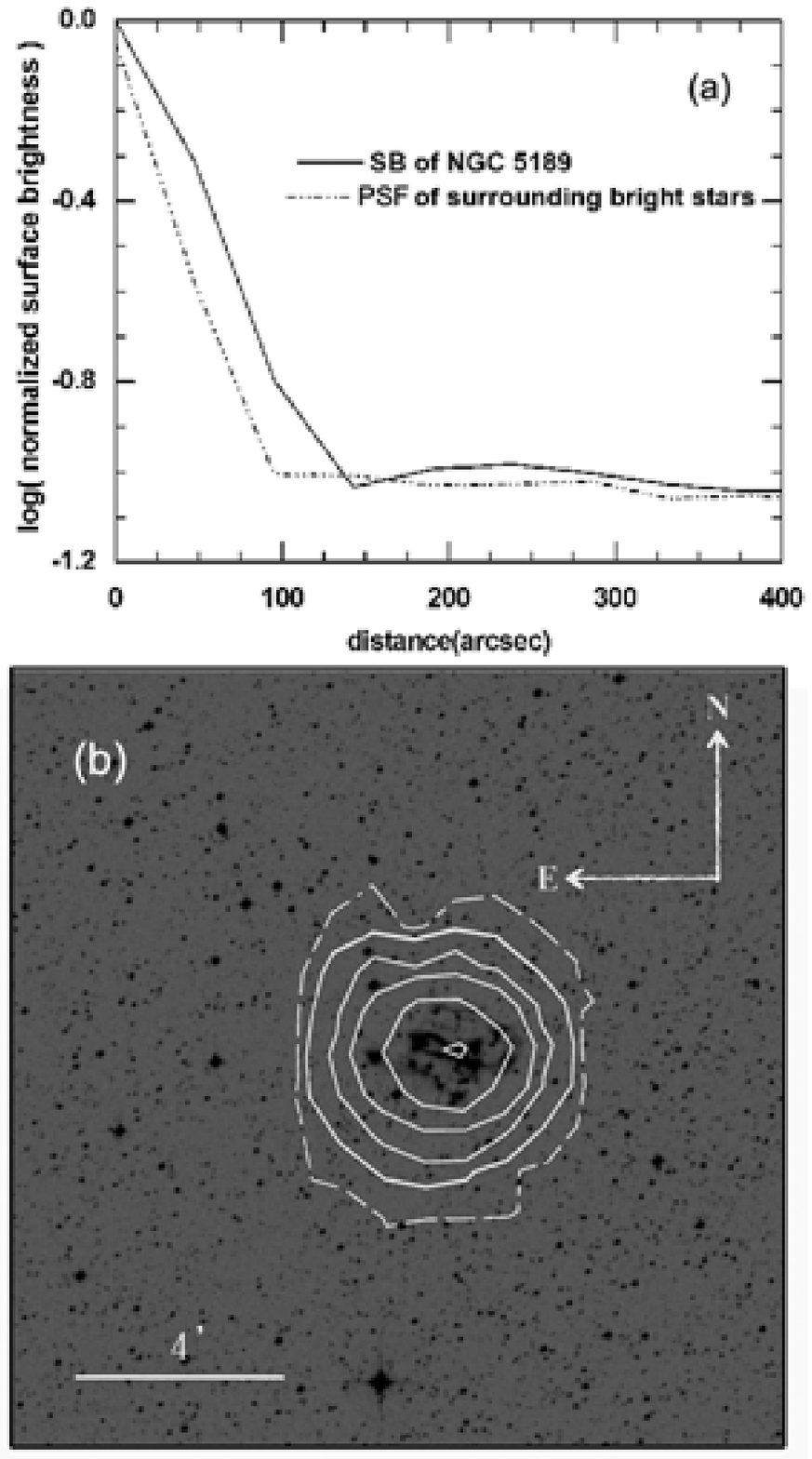}
      \caption{(a) Normalized SB profile of the halo
for NGC 5189 (solid line) and the PSF of nearby stars with similar
brightness (dotted line) along P.A. = 90 $^\circ$. (b) Contours of
the SHASSA imaging of the PN overlaid on the DSS II red plate
image with a spatial extension of 18.6$'$ $\times$
18.6$'$. The lowest level dotted contour corresponds to 1.83
$\times$ 10$^{-15}$ ergs $cm^{-2}$ $s^{-1}$ $arcsec^{-2}$, 3
$\sigma$ above the background. The other solid contours are 4.89
$\times$ 10$^{-15}$ (8 $\sigma$ above the background), 1.27
$\times$ 10$^{-14}$, 3.31 $\times$ 10$^{-14}$, 8.6 $\times$
10$^{-14}$, and 2.23 $\times$ 10$^{-13}$ ergs $cm^{-2}$ $s^{-1}$
$arcsec^{-2}$, respectively.}
         \label{Figure 9}
   \end{figure}

\begin{figure}
   \centering
   \includegraphics[width=10cm]{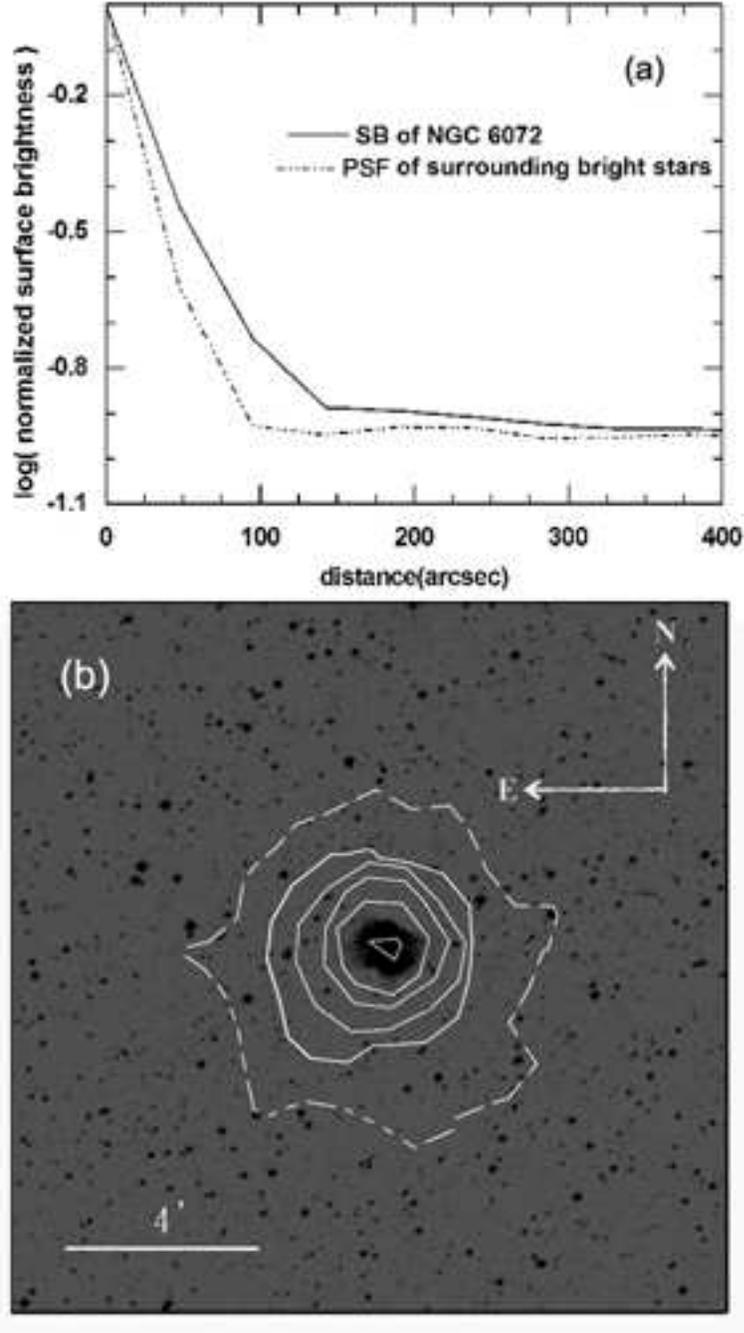}
      \caption{(a) The normalized SB profile of
the halo of NGC 6072  along P.A. = 90 $^\circ$ (solid line) and
the surrounding stars' PSF profile with similar brightness (dotted
line). (b) Contour plot of the SHASSA H$\alpha$ imaging overlaid
on a 18.6$'$ $\times$ 18.6$'$ DSS II red plate image. The
lowest level dotted contour corresponds to 1.05 $\times$
10$^{-15}$ ergs $cm^{-2}$ $s^{-1}$ $arcsec^{-2}$, 3 $\sigma$ above
the background. The other solid contours are 2.8 $\times$
10$^{-15}$ (8 $\sigma$ above the background), 6.73 $\times$
10$^{-15}$, 1.62 $\times$ 10$^{-14}$, 3.88 $\times$ 10$^{-14}$,
and 9.3 $\times$ 10$^{-14}$ ergs $cm^{-2}$ $s^{-1}$ $arcsec^{-2}$,
respectively.}
         \label{Figure 10}
   \end{figure}

\begin{figure}
   \centering
   \includegraphics[width=10cm]{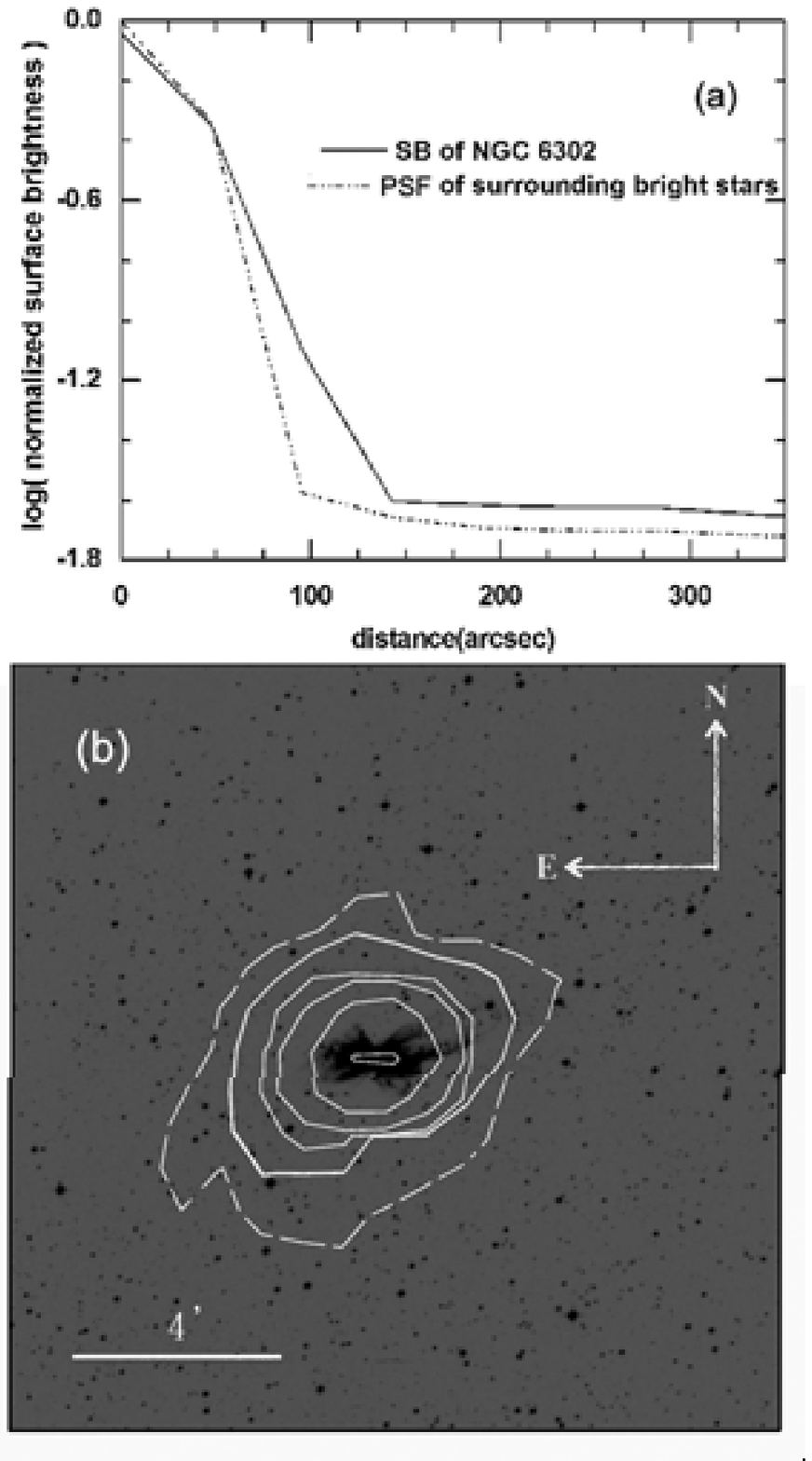}
      \caption{(a) Normalized SB profile of the
halo for NGC 6302  along P.A. = 0 $^\circ$ (solid line) and the
PSF of nearby stars with similar brightness (dotted line). (b)
Contours of the SHASSA imaging of the PN overlaid on the DSS II
red plate image with a spatial extension of 18.6$'$ $\times$
18.6$'$. The lowest level dotted contour corresponds to 3.02
$\times$ 10$^{-15}$ ergs $cm^{-2}$ $s^{-1}$ $arcsec^{-2}$, 3
$\sigma$ above the background. The other solid contours are 8.06
$\times$ 10$^{-15}$ (8 $\sigma$ above the background), 2.55
$\times$ 10$^{-14}$, 8.05 $\times$ 10$^{-14}$, 2.54 $\times$
10$^{-13}$, and 8.04 $\times$ 10$^{-13}$ ergs $cm^{-2}$ $s^{-1}$
$arcsec^{-2}$, respectively.}
         \label{Figure 11}
   \end{figure}

\end{document}